\documentclass[prl,twocolumn,showpacs,preprintnumbers,amsmath,amssymb]{revtex4}
\usepackage{epsfig}
\usepackage{wasysym}
\usepackage{amssymb}
\usepackage{slashed}
\usepackage{physics}
\usepackage{color}

\begin{document}

\newcommand{\bvec}[1]{\mbox{\boldmath ${#1}$}}
\title{Pure spin-3/2 propagator for use in particle and nuclear physics}
\author{J. Kristiano}
\author{S. Clymton}
\author{T. Mart}\email[]{terry.mart@sci.ui.ac.id}
\affiliation{Departemen Fisika, FMIPA, Universitas Indonesia, Depok 16424, 
  Indonesia}
\date{\today}
\begin{abstract}
  We propose the use of pure spin-3/2 propagator in the 
  $(3/2,0) \oplus (0,3/2)$ representation in particle and nuclear
  physics. To formulate the propagator in a covariant form we use
  the antisymmetric tensor spinor representation and we consider 
  the $\Delta$ resonance contribution to the elastic $\pi N$ scattering
  as an example. We find that the use
  of conventional gauge invariant interaction Lagrangian leads
  to a problem; the obtained scattering amplitude does not exhibit 
  the resonance behavior. To overcome this problem we modify the
  interaction by adding a momentum dependence. 
  As in the case of Rarita-Schwinger we find that a perfect resonance 
  description could be obtained in the pure spin-3/2 formulation only if 
  hadronic form factors were considered in the interactions. 
\end{abstract}
\pacs{11.10.Ef, 11.15.2q, 14.20.Gk, 13.75.Gx
}

\maketitle

For decades  the most commonly used propagator for 
spin-3/2 particles (e.g., the $N$ and $\Delta$ resonances)  
in particle and nuclear physics 
is the Rarita-Schwinger (RS) one \cite{rarita}, although it 
is also well known that the RS propagator has an intrinsic and
long-standing problem \cite{Williams:1985zz,Benmerrouche:1989uc}; 
it contains the unphysical extra degrees
of freedom (DOF) or lower spin background. To be more precise, 
let us begin with the RS field that is formed by the tensor product of  a vector and 
a Dirac field represented by $(1/2,1/2)$ and $(1/2,0) \oplus (0,1/2)$, respectively. 
This tensor product yields \cite{Weinberg:1995mt}
\begin{eqnarray}
&&
{\textstyle 
\left(\frac{1}{2},\frac{1}{2} \right)\otimes\left[ \left(\frac{1}{2},0 \right) 
\oplus \left(0,\frac{1}{2} \right) \right] }
\nonumber\\
&&
{\textstyle = 
 \left(1,\frac{1}{2} \right) 
\oplus \left(\frac{1}{2},1 \right) 
\oplus  \left(\frac{1}{2},0 \right) 
\oplus \left(0,\frac{1}{2} \right) ,}
\end{eqnarray}
which shows that the RS field consists of two fields; 
the $(1,1/2) \oplus (1/2,1)$ and the Dirac field. 
The orthogonality relation can be used to eliminate the Dirac field. 
The $(1,1/2) \oplus (1/2,1)$ is, however, still not free from the Dirac background.

So far, the popular choice for the 
interaction Lagrangian is given, e.g., in 
Eq.~(16) of Ref.~\cite{pascalutsa_1999}, which contains 
the so-called off-shell parameter. This parameter
is required to maintain the invariance of the RS Lagrangian
under point transformations \cite{Haberzettl:1998rw}.
In the phenomenological study of nuclear physics, however,
the physical meaning of the off-shell parameter raised a serious
problem, i.e., the coupling constants could heavily depend on 
the off-shell parameter \cite{Feuster:1996ww} and the $\Delta$ contribution 
for the Compton amplitude is obscured by this 
parameter \cite{Pascalutsa:2002pi}. There is also some infamous 
fundamental problem regarding the interaction with the electromagnetic 
field, i.e., the Johnson-Sudarshan~\cite{Johnson-Sudarshan} and 
Velo-Zwanziger~\cite{Velo-Zwanziger} problems. 
The origin of these problems comes from the unphysical degree of freedom 
that arises in the interaction for whatever choice we make for 
the off-shell parameter.

Furthermore, it is shown in 
Ref.~\cite{pascalutsa_1999} that this 
interaction does not posses any local symmetry of RS field and, 
as a consequence, it violates the constraints for reducing 
the number of independent components of the field to the correct value
and involves the unphysical lower-spin DOF. 
The pathologies of this interaction can be removed by 
introducing the gauge-invariant (GI) or consistent interaction to 
decouple the unphysical spin-1/2 background from the $\Delta$-exchanges 
amplitude~\cite{pascalutsa_1999}. 

Nevertheless, for the practical use of spin-3/2 propagators and 
couplings, e.g., in
the isobar or coupled-channels model for meson-nucleon scattering or 
meson induced reaction, 
a pure spin-3/2 propagator is strongly desired.
In particle physics there are 7 nucleon and 5 $\Delta$ resonances
with spins 3/2 \cite{pdg}. These resonances are also intensively used in
nuclear physics. Obviously, a solid formulation of the spin-3/2
amplitude is inevitable in these sectors. 

From the theoretical point of view, a particle with spin 3/2 can 
be represented by the pure 
spin-3/2 representation, $(3/2,0) \oplus (0,3/2)$, which 
is free from the spin-1/2 field. 
However, in the $(3/2,0) \oplus (0,3/2)$ representation 
the formulation involves an 8-dimensional
spinor because the spin-3/2 operator is represented by $4 \times 4$ 
matrices. Such an 8-dimensional spinor representation is not 
the popular choice because it cannot be written in a covariant form.
As a consequence it is hard to construct the corresponding interaction
Lagrangian. 

Fortunately, Acosta {\it et al.} \cite{DelgadoAcosta:2015ypa} 
show that the components of the 8-dimensional spinor can be embedded 
into a totally antisymmetric tensor of second rank. The representation 
is called the antisymmetric tensor spinor (ATS) representation, 
which is formed by the tensor product of an antisymmetric tensor and 
the Dirac field. The pure spin-3/2 representation is projected from 
the ATS representation. 

ATS is formed by a tensor product of an antisymmetric tensor field 
and the Dirac field, which is represented by $(1,0) \oplus (0,1)$ 
and $(1/2,0) \oplus (0,1/2)$, respectively. 
This tensor product may be expressed as \cite{DelgadoAcosta:2015ypa}
\begin{eqnarray}
&& {\textstyle
\left[ \left(1,0 \right) \oplus \left(0,1 \right) \right] 
\otimes\left[ \left(\frac{1}{2},0 \right) \oplus \left(0,\frac{1}{2} \right) 
\right] ~=~}
\nonumber\\ 
& & {\textstyle
\left[ \left(\frac{3}{2},0 \right) \oplus \left(0,\frac{3}{2} 
\right) \right] \oplus \left[ \left(1,\frac{1}{2} \right) \oplus 
\left(\frac{1}{2},1 \right) \right]}
\nonumber\\ && 
{\textstyle
 \oplus \left[ \left(\frac{1}{2},0 \right) \oplus \left(0,\frac{1}{2} 
\right) \right] .}
\end{eqnarray}
Thus, the ATS representation consists of two fields, i.e., the 
$(3/2,0) \oplus (0,3/2)$ and the RS field. 
The RS field can be removed from the ATS by operating 
the projection operator. To this end, one can define the Casimir operator
$F = {\textstyle\frac{1}{4}} J_{\mu \nu} J^{\mu \nu}$, where $J^{\mu\nu}$ is 
the angular momentum operator for the representation. The eigenstate equation of 
the Casimir operator for the field (a,b) reads
\begin{equation}
F \ket{(a,b)} = C(a,b) \ket{(a,b)},
\end{equation}
where $C(a,b) = a(a+1) + b(b+1)$. The $(1,1/2) \oplus (1/2,1)$ and 
$(1/2,0) \oplus (0,1/2)$ fields are removed from the ATS by the projection operator
\begin{equation}
\mathcal{P} = \frac{[F - C(1,{\textstyle\frac{1}{2}})] [(F - C({\textstyle\frac{1}{2}},0)]}{[C({\textstyle\frac{3}{2}},0) - C(1,{\textstyle\frac{1}{2}})] [C({\textstyle\frac{3}{2}},0) - C({\textstyle\frac{1}{2}},0)]} ~.
\end{equation}
Reference \cite{DelgadoAcosta:2015ypa} shows that the projection operator is equal to
\begin{equation}
\mathcal{P}_{\alpha \beta \gamma \delta} = {\textstyle\frac{1}{8}} \left( \sigma_{\alpha \beta} \sigma_{\gamma \delta} + \sigma_{\gamma \delta} \sigma_{\alpha \beta} \right) - {\textstyle\frac{1}{12}} \sigma_{\alpha \beta} \sigma_{\gamma \delta} \, ,
\end{equation}
where
\begin{equation}
\sigma_{\alpha \beta} = {\textstyle\frac{i}{2}}\left[ \gamma_\alpha, \gamma_\beta \right]\, .
\end{equation}
This projection operator ensures that the ATS has  only the 
$(3/2,0) \oplus (0,3/2)$ representation. The pure spin-3/2 ATS is obtained by 
operating a pure spin-3/2 projection operator to the GI RS spinor, i.e.,
\cite{DelgadoAcosta:2015ypa}
\begin{equation}
w^{\mu \nu}({\bf p}, \lambda) = 2{\mathcal{P}^{\mu \nu}}_{\alpha \beta} U^{\alpha \beta}({\bf p}, \lambda),
\end{equation}
where $\lambda = -3/2, -1/2, +1/2, +3/2$ are the $z$-components of the 
spin-3/2 operator eigenvalues and $U^{\alpha \beta}({\bf p}, \lambda)$ is 
the GI RS spinor, given by
\begin{equation}
U^{\alpha \beta}({\bf p}, \lambda) = \frac{1}{2m} \left[ p^\alpha \mathcal{U}^\beta({\bf p}, \lambda) - p^\beta \mathcal{U}^\alpha({\bf p}, \lambda)  \right],
\label{eq:U_ab}
\end{equation}
with $\mathcal{U}^\alpha({\bf p}, \lambda)$ the RS vector-spinor. 

Up to the normalization constant $(2m)^{-1}$ the GI 
RS spinor $U^{\alpha \beta}({\bf p}, \lambda)$ given in
Eq.~(\ref{eq:U_ab}) is identical to the GI RS field tensor $\Delta^{\mu \nu} = \partial^\mu 
\Delta^\nu - \partial^\nu \Delta^\mu$ given in 
Ref.~\cite{pascalutsa_1999}. 
Therefore, the ATS representation differs from the GI RS 
field tensor in the projection operator. Note that this projection operator 
is different from the common projection operator in RS field. This projection 
operator projects out the $(3/2,0) \oplus (0,3/2)$ field from the ATS representation, 
instead of the $(1,1/2) \oplus (1/2,1)$ field. The propagator of pure spin-3/2 
representation reads
\begin{equation}
S_{\alpha \beta \gamma \delta}(p) = \frac{\Delta_{\alpha \beta \gamma \delta}(p)}{p^2 - m^2 + i\epsilon},
\end{equation}
where
\begin{equation}
\Delta_{\alpha \beta \gamma \delta}(p) = \left( \frac{p^2}{m^2} \right) \mathcal{P}_{\alpha \beta \gamma \delta} - \left( \frac{p^2-m^2}{m^2} \right) 1_{\alpha \beta \gamma \delta},
\label{eq:purepropagator}
\end{equation}
and $1_{\alpha \beta \gamma \delta}$ is the identity in ATS space with
\begin{equation}
1_{\alpha \beta \gamma \delta} = {\textstyle\frac{1}{2}} \left( g_{\alpha \gamma}g_{\beta \delta} - g_{\alpha \delta}g_{\beta \gamma} \right)1_{4 \times 4}.
\end{equation}
Based on the orthogonality relation for the projection operator 
$\gamma^\mu \mathcal{P}_{\mu \nu \rho \sigma} = 0$, one can prove that the pure 
spin-3/2 spinor satisfies  $\gamma_\mu w^{\mu \nu}({\bf p}, \lambda) = 0$. 
This relation reduces the number of DOF of the ATS representation, i.e.,
$6 \times 4 = 24$, by  $4 \times 4 = 16$. As expected, the pure spin-3/2 field 
in the ATS representation has $24-16=8$ DOF. 

The free Lagrangian for pure spin-3/2 field in the ATS representation is given 
by \cite{DelgadoAcosta:2015ypa}
\begin{equation}
\mathcal{L} = (\partial^\mu \Psi^{\alpha \beta}) \Gamma_{\mu \nu \alpha \beta \gamma \delta} (\partial^\nu \Psi^{\gamma \delta}) - m^2 \Psi^{\mu \nu} \Psi_{\mu \nu} \, ,
\end{equation}
where $\Gamma_{\mu \nu \alpha \beta \gamma \delta} = 	4 g^{\sigma \rho} \mathcal{P}_{\alpha \beta \rho \mu} \mathcal{P}_{\sigma \nu \gamma \delta}$ and $\Psi^{\mu \nu}$ 
is the $(3/2,0) \oplus (0,3/2)$ field. The kinetic term of the Lagrangian is invariant 
under the following gauge transformation
\begin{equation}
\Psi^{\mu \nu} \rightarrow \Psi^{\mu \nu} + \xi^{\mu \nu} \, ,
\label{eq:gaugetrans}
\end{equation}
where $\xi^{\mu \nu}$ is an antisymmetric tensor containing the gamma matrices
\begin{equation}
\xi^{\mu \nu} = \gamma^\mu \partial^\nu \xi - \gamma^\nu \partial^\mu \xi \, .
\end{equation}

\begin{figure}[t]
  \begin{center}
    \leavevmode
    \epsfig{figure=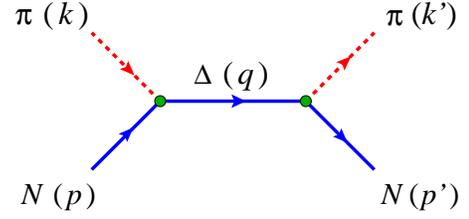,width=60mm}
    \caption{(Color online) Feynman diagram for the elastic $\pi N$ 
      scattering with a $\Delta$ resonance in the intermediate state.
      }
   \label{fig:feynman} 
  \end{center}
\end{figure}

Let us consider the $\pi N\to\pi N$ scattering with a $\Delta$ resonance
in the intermediate state as an example. The corresponding Feynman diagram is depicted
in Fig.~\ref{fig:feynman}, where the momenta of the involved particles are
also shown for our convention.
The popular choice for $\pi N \Delta$ Lagrangian interaction 
is \cite{pascalutsa_1999}
\begin{equation}
\mathcal{L}_{\pi N \Delta} = \left( \frac{g_{\pi N \Delta}}{m_\pi} \right) 
\bar{\Delta}^\mu \Theta_{\mu \nu}(z) N \partial^\nu \pi + {\rm H.c.},
\end{equation}
where $\Theta_{\mu \nu}(z)$ is given by
\begin{equation}
\Theta_{\mu \nu}(z) = g_{\mu \nu} - \left(  z+{\textstyle\frac{1}{2}} \right) \gamma_\mu \gamma_\nu,
\label{eq:theta_mu_nu}
\end{equation}
the constant $z$ is arbitrary and conventionally called the off-shell 
parameter. Furthermore, $\Delta^\mu$, $N$, and $\pi$ in 
Eq.~(\ref{eq:theta_mu_nu}) denote the 
$\Delta$-baryon vector-spinor, nucleon spinor, and pion field respectively. 
As stated before this Lagrangian does not posses any local symmetries of 
the RS field, and as a consequence it induces the unphysical lower-spin 
DOF, which is called spin 1/2 background \cite{pascalutsa_1999}. 
To decouple this unphysical background from the $\Delta$-exchange amplitude 
Pascalutsa and Timmermans introduced a GI interaction which is given by \cite{pascalutsa_1999}
\begin{equation}
\mathcal{L}_{\pi N \Delta} = \left( \frac{g_{\pi N \Delta}} {m_\pi m_\Delta} \right) 
\bar{N} \gamma_5 \gamma_\mu \tilde{\Delta}^{\mu \nu}\partial_\nu \pi + {\rm H.c.},
\end{equation}
where $\tilde{\Delta}^{\mu \nu}$ is the dual tensor of GI 
RS field tensor $\Delta^{\mu \nu}$ defined as
\begin{equation}
\Delta_{\mu \nu} = \partial_\mu \Delta_\nu - \partial_\nu \Delta_\mu .
\end{equation}
This GI interaction yields the $\Delta$-exchange amplitude
\begin{equation}
\Gamma^\mu (k') S_{\mu \nu}(q) \Gamma^\nu (k) = \frac{(g_{\pi N \Delta}/m_\pi)^2}{\slashed{q} 
- m_\Delta} \frac{q^2}{m_\Delta^2} P_{\mu \nu}^{(3/2)}(q) k'^\mu k^\nu,
\label{eq:GI_interaction}
\end{equation}
with $P_{\mu \nu}^{(3/2)}$ the spin-3/2 projection operator in the RS field given by
\begin{equation}
P_{\mu \nu}^{(3/2)}(q) = g_{\mu\nu} - {\textstyle\frac{1}{3}} \gamma_\mu \gamma_\nu - \frac{1}{3q^2} (\slashed{q} \gamma_\mu q_\nu + q_\mu \gamma_\nu \slashed{q}).
\end{equation}

Analogous to the GI interaction, one can construct the 
$\pi N \Delta$ interaction by changing the GI RS field 
tensor to the $(3/2,0) \oplus (0,3/2)$ representation
\begin{equation}
\mathcal{L}_{\pi N \Delta} = g_{\pi N \Delta}
\bar{N} \gamma_5 \gamma_\mu \tilde{\Psi}^{\mu \nu}\partial_\nu \pi + {\rm H.c.},
\label{eq:L_piND}
\end{equation}
where $\Psi^{\mu\nu}$ is the $(3/2,0) \oplus (0,3/2)$ field tensor 
and $\tilde{\Psi}^{\mu \nu}$ is its dual tensor. By using the vertex 
factor 
\begin{equation}
\Gamma_{\mu\nu}(k) = g_{\pi N \Delta} \gamma_5 \gamma_\mu k_\nu ,
\end{equation}
the corresponding $\Delta$-exchange amplitude becomes 
$\Gamma_{\mu\nu}(k') \tilde{S}^{\mu \nu \rho \sigma}(q) \Gamma_{\rho \sigma}(k)$, 
where $\tilde{S}^{\mu \nu \rho \sigma}$ is defined by 
\begin{eqnarray}
\Gamma_{\mu\nu}(k') \tilde{S}^{\mu \nu \rho \sigma}(q) \Gamma_{\rho \sigma}(k) &=& 
{\textstyle\frac{1}{4}} g_{\pi N \Delta}^2 \epsilon^{\mu \nu \alpha \beta} 
\epsilon^{\rho \sigma \kappa \lambda} \gamma_5 \gamma_\mu \nonumber\\
&\times& S_{\alpha \beta \kappa \lambda}(q) \gamma_\rho \gamma_5 k'_\nu k_\sigma \, .
\label{eq:delta_exchange}
\end{eqnarray}
By evaluating Eq.~(\ref{eq:delta_exchange}) we find that 
the nonvanishing $\Delta$-exchange amplitude originates
from the contraction with $1_{\alpha \beta \kappa \lambda}$, 
since the contraction with $\mathcal{P}_{\alpha \beta \kappa \lambda}$ 
vanishes due to the orthogonality relation 
$\gamma^\alpha \mathcal{P}_{\alpha \beta \kappa\lambda} = 0$ 
and $\tilde{\sigma}^{\mu \nu} = - \gamma_5 \sigma^{\mu \nu}$. 
After some calculations we obtain the $\Delta$-exchange amplitude in
the form of
\begin{eqnarray}
\Gamma_{\mu\nu}(k') \tilde{S}^{\mu \nu \rho \sigma}(q) \Gamma_{\rho \sigma} (k)
&=& \frac{g_{\pi N \Delta}^2 \left( q^2-m_\Delta^2 \right)}{m_\Delta^2 
(q^2 - m_\Delta^2 + i\epsilon)}\nonumber\\
&\times& \left(  g^{\nu \sigma} + 
{\textstyle\frac{1}{2}} \gamma^\nu \gamma^\sigma \right) k'_\nu k_\sigma \, . ~~
\label{eq:gammap_mu_nu}
\end{eqnarray}
Based on Eq.~(\ref{eq:gammap_mu_nu}) we can conclude that this 
$\Delta$-exchange amplitude cannot describe the contribution of 
a resonance, since at the resonance pole ($q^2 = m_\Delta^2$) 
the amplitude is equal 
to zero instead of being maximum. As a consequence, the Lagrangian 
given in Eq.~(\ref{eq:L_piND}) cannot be used for calculating
the resonance contribution.

The problem originates from the GI interaction Lagrangian given in
Eq.~(\ref{eq:L_piND}), since the 
contraction between gamma matrices and the pure spin-3/2 field tensor 
vanishes. To overcome this problem one can modify the interaction Lagrangian 
by replacing the gamma matrix with a partial derivative, i.e.,
\begin{equation}
\mathcal{L}_{\pi N \Delta} = \left( \frac{g_{\pi N \Delta}}{m_\Delta} \right) 
\bar{N} \gamma_5 \partial^\mu \Psi_{\mu \nu}\partial^\nu \pi + {\rm H.c.}
\label{eq:LPiNDpartial}
\end{equation}
By using the vertex factor
\begin{equation}
\Gamma^{\mu \nu}(k) = \left( \frac{g_{\pi N \Delta}}{m_\Delta} \right) \gamma_5 q^\mu k^\nu ,
\end{equation}
the $\Delta$-exchange amplitude corresponding to this interaction reads 
\begin{eqnarray}
&&\Gamma^{\mu \nu}(k') S_{\mu \nu \rho \sigma}(q) \Gamma^{\rho \sigma}(k) \nonumber\\
&=& \left( \frac{g_{\pi N \Delta}}{m_\Delta} \right)^2 \gamma_5 q^\mu 
 S_{\mu \nu \rho \sigma}(q) \gamma_5 q^\rho k'^\nu k^\sigma.
\end{eqnarray}
After some calculations we obtain
\begin{eqnarray}
&&\Gamma^{\mu \nu}(k') S_{\mu \nu \rho \sigma}(q) \Gamma^{\rho \sigma}(k) \nonumber\\
&=& \frac{g_{\pi N \Delta}^2 k'^\nu k^\sigma}{q^2 - m_\Delta^2 + i\epsilon} 
\left[ \frac{q^4}{4m_\Delta^4} P^{(3/2)}_{\nu \sigma}(q)  \right.
\nonumber\\
&-& \left. 
\left( \frac{q^2 - m_\Delta^2}{2m_\Delta^4} \right) \left(  q^2 g_{\nu \sigma} - q_\nu q_\sigma \right) \right],~~~~
\label{eq:amplitude_pure}
\end{eqnarray}
which differs from the result of GI interaction given by Eq.~(\ref{eq:GI_interaction}) by the second term. 
However, this result is very interesting, because at
the resonance pole, i.e., $q^2 = m_\Delta^2$, the second term vanishes and 
the $\Delta$-exchange amplitude is proportional to RS spin-3/2 
projection operator.

For future consideration we need to point out here that the
GI electromagnetic interaction reads \cite{pascalutsa_1999}
\begin{eqnarray}
\mathcal{L}_{\gamma N \Delta} &=& e \bar{N} \left( g_1 \tilde{\Delta}_{\mu \nu} + 
g_2 \gamma_5 \Delta_{\mu \nu} + g_3 \gamma_\mu \gamma^\rho \tilde{\Delta}_{\rho \nu}  \right.
\nonumber\\
&+& \left.
g_4 \gamma_5 \gamma_\mu \gamma^\rho \Delta_{\rho \nu} \right) F^{\mu \nu} + {\rm H.c.} ,
\end{eqnarray}
whereas the non-GI interactions are \cite{pascalutsa_1999}
\begin{eqnarray}
\mathcal{L}^{(1)}_{\gamma N \Delta} &=& \frac{ie G_1}{2m} \bar{\Delta}^\rho \Theta_{\rho \mu}(z_1) \gamma_\nu 
\gamma_5 N F^{\mu \nu} + {\rm H.c.} ,\nonumber\\
\mathcal{L}^{(2)}_{\gamma N \Delta} &=& -\frac{e G_2}{(2m)^2} \bar{\Delta}^\rho \Theta_{\rho \mu}(z_2) \gamma_5 
\partial_\nu N F^{\mu \nu} + {\rm H.c.} ,
\end{eqnarray}
where $z_1$ and $z_2$ are the off-shell parameters.
Interestingly, the pure spin-3/2 interaction is given by 
\begin{equation}
\mathcal{L}_{\gamma N \Delta} = e \bar{N} \left( f_1 \Psi_{\mu \nu} + f_2 \gamma_\mu 
\partial^\rho \Psi_{\rho \nu}  \right) F^{\mu \nu} + {\rm H.c.} ,
\end{equation}
which differs from the GI one  by the number of coupling constants, i.e.,
the pure spin-3/2 representation has only two couplings 
because $\tilde{\Psi}_{\mu \nu} = -\gamma_5 \Psi_{\mu \nu} $. Thus, the 
interaction of photon with pure spin-3/2 resonance has the same number 
of couplings as in the case of non-GI model.

To visualize the behavior of the pure spin-3/2 propagator 
we will compare the contributions of the spin-3/2
$\Delta(1232)$ resonance 
amplitudes obtained from the pure spin-3/2 propagator and 
from the Rarita-Schwinger one 
to the total cross section of elastic $\pi N$ scattering. 
To this end we 
include the resonance width $\Gamma$ in the resonance 
propagator by replacing $i\epsilon\to i\Gamma m_\Delta$ and 
write the scattering amplitude in the form of
\begin{eqnarray}
\mathcal{M}&=&\bar{u}(p',s')(A+B\slashed{Q})u(p,s) ,
\label{eq:amplitude_M}
\end{eqnarray}
with $Q=(k+k')/2$. For the RS propagator with GI interaction we obtain
\begin{eqnarray}
A &=& G \left\lbrace m_N\left(3k' \cdot k - 2 p \cdot k - m_\pi^2 - {2}q \cdot k' q \cdot k/{q^2} \right)\right.\nonumber\\
& & \left. + m_\Delta\left(3k'\cdot k-2p\cdot k-m_\pi^2 -{2}m_\pi^2 q\cdot Q/{q^2}\right)\right\rbrace,~~~~\\
B &=& G \left\lbrace 3k' \cdot k -m_\pi^2 +2m_N^2 -{2} q \cdot k' q \cdot k/{q^2} \right.\nonumber\\
& & \left. +q\cdot (k'-k) + 2m_\Delta m_N \left( 1 - {q \cdot Q}/{q^2}\right)\right\rbrace ,
\end{eqnarray}
with 
\begin{eqnarray}
G &=& {q^2\,g_{\pi N\Delta}^2}/[{3m_\pi^2 m_\Delta^2(q^2-m_\Delta^2 +i\Gamma m_\Delta)}] .
\end{eqnarray}

In the case of pure spin-3/2 propagator we have
\begin{eqnarray}
A &=& G\Bigl[ ({q^4}/{12m_\Delta^4}) \Bigl(3k' \cdot k - m_\pi^2 - 2 p \cdot k \nonumber\\
&& - 2 m_\pi^2\, q \cdot Q/{q^2} \Bigr)  - \left\{ (q^2 - m_\Delta^2)/2 m_\Delta^4\right\}\nonumber\\
& & \times (q^2 k' \cdot k - q \cdot k q \cdot k')\Bigr] ,~~~\\
B &=& \left({q^4 m_N G}/{6m_\Delta^4}\right) \left( 1 - {q \cdot Q}/{q^2}  \right) ,
\end{eqnarray}
with 
\begin{eqnarray}
G &=& {g_{\pi N\Delta}^2}/[q^2 - m_\Delta^2 + i\Gamma m_\Delta] .
\end{eqnarray}
The cross section can be obtained from the scattering amplitude $\mathcal{M}$
given by Eq.~(\ref{eq:amplitude_M}) by means of the standard
method \cite{bjorken}.
Since the analytic forms of the two amplitudes are completely different
we have to use different coupling constants in order  to produce comparable results.
We believe that this should not raise a problem since the coupling
constants are usually fitted to reproduce the experimental data.

\begin{figure}[t]
  \begin{center}
    \leavevmode
    \epsfig{figure=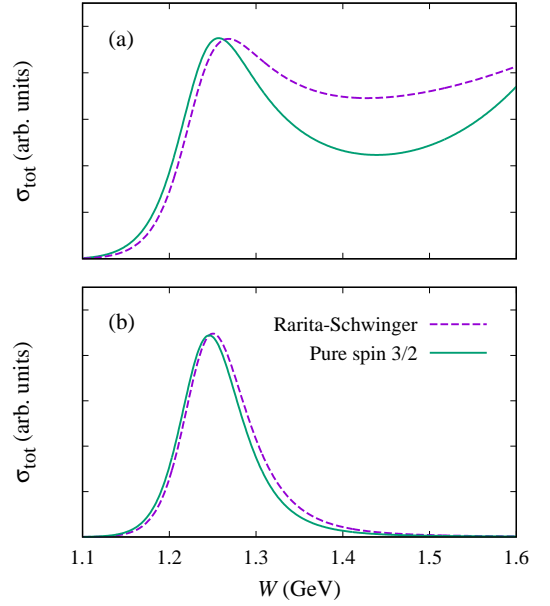,width=70mm}
    \caption{(Color online) Contribution of the $\Delta(1232)$ resonance 
      to the $\pi N\to \pi N$ scattering total cross section in arbitrary
      units (arb. units) according to the Rarita-Schwinger with GI interaction 
      and the pure spin-3/2 models as a function of total c.m. energy $W$.
      Panels (a) and (b) show the contributions if the hadronic form factors
      are excluded and included, respectively.
      Note that for the sake of comparison the two models do not use
      the same value of coupling constant.
      }
   \label{fig:contribution} 
  \end{center}
\end{figure}

The original contributions of both models are depicted in panel (a) of
Fig.~\ref{fig:contribution}, where we can see the resonance behavior
centered around $W\approx 1.25$ GeV, followed by the background 
phenomenon originating from momentum dependence of the numerator of 
Eq.~(\ref{eq:amplitude_pure}) indicated by
a smoothly increasing cross section for $W\gtrsim 1.40$ GeV. Note that
this background shifts the resonance peak from the original position
at 1.232 GeV to higher energy region. It is also 
obvious from panel (a) of Fig.~\ref{fig:contribution} that
the background obtained from the pure spin-3/2 model is significantly smaller than
that of the RS model at $W\approx 1.40$ GeV. This phenomenon originates from the
second term in the square bracket 
of Eq.~(\ref{eq:amplitude_pure}). Above this energy 
point the first
term of Eq.~(\ref{eq:amplitude_pure}) starts to become dominant, since $q^4=W^4$, 
and the total contribution starts to diverge.

The large background contribution is natural in the covariant Feynman diagrammatic
approach. Alternatively, this could be also interpreted as the contribution of a Z-diagram
\cite{Jaroszewicz:1990mx}, i.e., the existence
of particle and antiparticle in the intermediate state, that is not considered 
in Fig.~\ref{fig:feynman}. The situation is quite different in the multipoles approach, where a relatively 
perfect resonance structure is parameterized by using the Breit-Wigner
function \cite{Mart:2017mwj}. However, in a covariant isobar model \cite{Clymton:2017nvp}
the large background 
contributions produced by a number of resonances in the model could disturb the 
nature of the resonance itself and increase the difficulty to fit the
experimental data. To suppress this undesired 
background one usually considers a hadronic form factor (HFF) in each of hadronic vertices 
of Fig.~\ref{fig:feynman}. A related discussion to this end can be found, e.g., in 
Refs.~\cite{Vrancx:2011qv,hbmf}. Nevertheless, here we have to emphasize that the
use of HFF is theoretically required to account for the fact that the 
nucleon is not a point-like object.
In the present analysis we use a dipole HFF in the form of \cite{hbmf}
\begin{equation}
F = {\Lambda^4}/[\Lambda^4+\left( q^2-m_\Delta^2 \right)^2] ,
\end{equation}
with the hadronic cutoff $\Lambda=0.5$ GeV. The choice seems to be trivial, but
at present it is solely intended for example. A detailed study for this purpose
can be addressed in the future. 

The result obtained 
after including this HFF
is shown in panel (b) of Fig.~\ref{fig:contribution}. Clearly, we obtain a perfect
resonance structure for both models and the fact that the RS structure is slightly shifted
to the right can be understood from its original contribution shown in panel (a)
of Fig.~\ref{fig:contribution}. Therefore, instead of its different formulation 
the pure spin-3/2 propagator also exhibits a common resonance structure as in the
conventional RS propagator.
This result also emphasizes our argument that in order to
obtain the natural properties of resonance the use of HFF is mandatory in the covariant 
Feynman diagrammatic approach.

Finally, one could also raise question about the consistency of the 
interaction Lagrangian given in Eq.~(\ref{eq:LPiNDpartial}). 
According to Ref.~\cite{Pascalutsa:2000kd}, an interaction is said to be 
consistent if the interaction Lagrangian has the same symmetry 
as in the free Lagrangian, i.e., the invariance under the same transformation 
as in the free Lagrangian. The Lagrangian given in Eq.~(\ref{eq:LPiNDpartial}) 
is not invariant under the gauge transformation given by Eq.~(\ref{eq:gaugetrans}). 
However, it is not difficult to construct a consistent Lagrangian for this
purpose. The general form of the interaction Lagrangian reads
\begin{equation}
\mathcal{L} = g \bar{J}_{\mu\nu} \Psi^{\mu \nu} + \mathrm{H.c.}
\end{equation}
The invariance of this interaction under the gauge transformation 
given in Eq.(\ref{eq:gaugetrans}) requires that 
\begin{equation}
\bar{J}_{\mu \nu} \xi^{\mu \nu} = 0,
\end{equation}
whereas $J_{\mu \nu}$ must not be a symmetric tensor because in general 
one does not expect that $J_{\mu \nu} \Psi^{\mu \nu}$ vanishes. 
One of the possible choices for the  consistent Lagrangian is 
\begin{equation}
\mathcal{L}_{\pi N \Delta} = \left( \frac{g_{\pi N \Delta}}{m_\Delta} 
\right) \bar{N} \gamma_5 \mathcal{P}_{\mu \nu \rho \sigma} \partial^\rho 
\Psi^{\mu\nu} \partial^\sigma \pi + \mathrm{H.c.}
\end{equation}
By using the vertex factor
\begin{equation}
\Gamma_{\mu \nu}(k) = \left( \frac{g_{\pi N \Delta}}{m_\Delta} \right) \gamma_5 \mathcal{P}_{\mu \nu \rho \sigma} q^\rho k^\sigma \, ,
\end{equation}
the corresponding $\Delta$-exchange amplitude for this interaction reads 
\begin{eqnarray}
&&\Gamma_{\mu \nu}(k') S^{\mu \nu \rho \sigma}(q) \Gamma_{\rho \sigma}(k) \nonumber\\
&=& \left( \frac{g_{\pi N \Delta}}{m_\Delta} \right)^2 q^\alpha k'^\beta \mathcal{P}_{\mu \nu \alpha \beta} S^{\mu \nu \rho \sigma}(q)  \mathcal{P}_{\rho \sigma \gamma \delta} q^\gamma k^\delta \, , ~~
\end{eqnarray}
With the idempotent relation of the projection operator 
$\mathcal{P}^{\mu \nu \alpha \beta} \mathcal{P}_{\alpha \beta \rho \sigma} = {\mathcal{P}^{\mu \nu}}_{\rho \sigma}$, 
one can prove the relation $\mathcal{P}^{\mu \nu \alpha \beta} \Delta_{\alpha \beta \rho \sigma} = {\mathcal{P}^{\mu \nu}}_{\rho \sigma}$. 
As a result, the transition amplitude becomes
\begin{eqnarray}
&&\Gamma_{\mu \nu}(k') S^{\mu \nu \rho \sigma}(q) \Gamma_{\rho \sigma}(k) \nonumber\\
&=& \frac{g_{\pi N \Delta}^2}{m_\Delta^2 (q^2 - m_\Delta^2 + i\epsilon)} 
\left[ q^\alpha k'^\beta \mathcal{P}_{\alpha \beta \gamma \delta} q^\gamma k^\delta \right] \nonumber\\
&=& \frac{g_{\pi N \Delta}^2}{q^2 - m_\Delta^2 + i\epsilon} 
\left[ \frac{q^2}{4m_\Delta^2} P^{(3/2)}_{\beta \delta} k'^\beta k^\delta \right]
\end{eqnarray}
Surprisingly, this transition amplitude contains only the RS spin-3/2 projection 
operator term. This transition amplitude differs from the result of
Pascalutsa and Timmermans \cite{pascalutsa_1999}  by the $(\slashed{q} + m_\Delta)$ factor.

In conclusion we have proposed the use of pure spin-3/2 propagator to describe the
properties of spin-3/2 particles in the study of particle and nuclear physics. We
used the ATS representation to describe the corresponding projection operator. We have
shown that in the ATS formalism we have to redefine the interaction Lagrangian, otherwise
the obtained scattering amplitude cannot display the resonance behavior. By calculating
its contribution to the elastic $\pi N$ scattering total cross section
we have shown that this pure spin-3/2 propagator also exhibits the natural properties 
of a resonance, as in the conventional RS one, if the hadronic form factors were
considered in its hadronic interactions.

The authors acknowledge the support from the PITTA Grant of Universitas Indonesia, 
under contract No. 689/UN2.R3.1/HKP.05.00/2017.

\end{document}